\documentstyle[prb,aps,multicol,epsfig,amssymb]{revtex}

\begin{document}
\draft

\title{Phase diagram of a random-anisotropy mixed-spin Ising model}
\author{A. P. Vieira, J. X. de Carvalho, and S. R. Salinas}
\address{Instituto de F\'{\i}sica, Universidade de S\~{a}o Paulo\\
Caixa Postal 66318 \\
05315-970, S\~{a}o Paulo, SP, Brazil}
\date{\today}
\maketitle

\begin{abstract}
We\ investigate the phase diagram of a mixed spin-$\frac{1}{2}$--spin-$1$ Ising
system in the presence of quenched disordered anisotropy. We carry out a
mean-field and a standard self-consistent Bethe--Peierls calculation.
Depending on the amount of disorder, there appear novel transition lines and
multicritical points. Also, we report some connections with a percolation
problem and an exact result in one dimension.
\end{abstract}
\pacs{05.50.+q, 64.60.Kw}
\ifpreprintsty
\else
\begin{multicols}{2}
\fi
\section{Introduction}

Apart from their relevance to the description of ferrimagnetic materials, 
mixed-spin models are interesting from a purely theoretical point of view,
being among the simplest models to exhibit tricritical behavior. So
they are 
especially convenient for studying the effects of inhomogeneities on 
the phase diagrams and the multicritical behavior of magnetic systems.
From a few exact {\cite{goncalves85,dasilva91}} and several approximate
\cite{zhang93,quadros94,buendia97,tucker99} calculations, we now have a 
good picture of the phase
diagrams of mixed spin-$\frac{1}{2}$--spin-$1$ Ising models in the presence
of a crystal field. Our aim in this work is to use this model Hamiltonian to
investigate the effects of disorder on the location of the transition lines
and the tricritical point. 

The mixed-spin Ising model is described as a two-sublattice system, with 
spin variables $\sigma =\pm 1$ and $S=0,\pm 1$, on the sites of sublattices 
$A$ and $B$, respectively.
Restricting the interactions to nearest-neighbors (belonging to different
sublattices) and single-ion terms, the most general spin Hamiltonian in even
spin space is written as 
\begin{equation}
H=-J\sum_{\left\langle i\in A,j\in B\right\rangle }\sigma
_{i}S_{j}+D\sum_{j\in B}S_{j}^{2},  \label{smp}
\end{equation}
where the first sum is over nearest-neighbor pairs, the second sum is over
the sites of sublattice $B$, and the parameter $J$ is assumed to be
positive (ferromagnetic exchange). For $D>0$, the 
crystal field favors the $S_{j}=0$ states; the
competition between anisotropy and ferromagnetic exchange terms leads to the
appearance of a tricritical point. It should be pointed out that one needs
three parameters to describe the even space of the better known spin-$1$
Blume-Emery-Griffiths (BEG) model. In the present case of a mixed-spin Ising
system, from the point of view of the calculations, the reduction of the
analysis to a two-parameter space is a particularly attractive feature.

There are exact calculations for the thermodynamic functions associated with
the model Hamiltonian given by Eq. (\ref{smp}) on a simple chain and on some
three-coordinated two-dimensional structures. On a honeycomb lattice, the
problem can be mapped onto a spin-$\frac{1}{2}$ Ising model on a triangular 
lattice, which does not display a tricritical point.\cite{goncalves85} This
mixed-spin model can also be exactly solved on a Bethe lattice 
\cite{dasilva91} (the deep interior of Cayley tree), leading to the same 
results of a recent cluster-variational calculation.\cite{tucker99} The 
results on a Bethe lattice with coordination $q$ indicate the absence of a 
tricritical point for $q<5$, as confirmed by Migdal--Kadanoff 
renormalization-group calculations.\cite{quadros94} In the 
infinite-coordination limit of the Bethe lattice, one regains the well-known 
results for the tricritical point
displayed by the Curie--Weiss (mean-field) version of the model. An earlier
approximate effective-field calculation \cite{kaneyoshi87} predicted a
tricritical point for $q\geq 4$, but this result has been recently
challenged.\cite{bobak97,delima00}

In order to analyze the effects of disorder, we consider the Hamiltonian 
\begin{equation}
H=-J\sum_{\left\langle i\in A,j\in B\right\rangle }\sigma
_{i}S_{j}+\sum_{j\in B}D_{j}S_{j}^{2},  \label{sphd}
\end{equation}
where $\left\{ D_{j}\right\} $ is a set of independent, identically
distributed random variables associated with the binary probability
distribution 
\begin{equation}
\wp (D_{j})=p\delta (D_{j})+(1-p)\delta (D_{j}-D).  \label{smprob}
\end{equation}
With this choice of disorder, and for $D>qJ$, the ground state can be mapped
onto a percolation problem in which the dilution affects the sites belonging
to only one of the sublattices (corresponding to spin $S=1$). This
association is easy to see if we note that a uniform crystal field $D>qJ$
leads to $S_{j}=0$ for all $j$, which breaks the connectivity between the
spin-$1/2$ variables. The presence of a randomly located distribution of $%
D=0 $ crystal fields recovers that connectivity and, for sufficiently high
values of $p$, leads to the formation of a percolating cluster. In the
rather artificial case of annealed disorder, on the honeycomb lattice, there
is also an exact solution \cite{goncalves92}\ for the thermodynamic
properties of the mixed-spin model described by Eqs. (\ref{sphd}) and (\ref
{smprob}). It is interesting to remark that this solution in the annealed
case reproduces the critical concentration of the percolation problem
associated with the ground state of the model with quenched (frozen)
disorder, which is equivalent to the usual percolation problem on the
triangular lattice. For the physically more relevant case of quenched
disorder, there are approximate calculations using an effective-field theory
with correlations,\cite{kaneyoshi88} which point to the (expected)
weakening of the tricritical behavior due to the presence of disorder.

In the present work, we first analyze the temperature--anisotropy ($T\times D
$) phase diagram of the Curie--Weiss version (mean-field limit) of the
spin Hamiltonian given by Eqs. (\ref{sphd}) and (\ref{smprob}). Depending on
the concentration $p$, there appear novel transition lines and multicritical
points. To include the effects of thermal fluctuations, we then resort to a
standard self-consistent Bethe--Peierls approximation (which is analogous,
in the case of the corresponding uniform model, to performing an exact
Bethe-lattice calculation). In section II, we present the Curie--Weiss
results. The Bethe--Peierls approximation is discussed in section III. In
the appendices, we report an exact solution in one-dimension as well as some
low-temperature expansions to supplement the numerical results from the
Bethe--Peierls approximation.

\section{Curie--Weiss version}

The Curie--Weiss version of the mixed-spin Ising model is given by the
Hamiltonian 
\begin{equation}
H=-\frac{2J}{N}\sum_{i\in A}\sigma _{i}\sum_{j\in B}S_{j}+\sum_{j\in
B}D_{j}S_{j}^{2},
\end{equation}
where the sums are over all sites belonging to each one of the sublattices.

For a given disorder configuration $\{D_{j}\}$, we calculate the partition
function by performing a partial trace over the set of spin variables $%
\{S_{j}\}$. In the thermodynamic limit, we use the saddle-point method and
average over disorder to obtain the free energy functional 
\ifpreprintsty
\begin{eqnarray}%
\Psi(\sigma)&=&-\frac{1}{2\beta}\left[\ln2-\frac{1}{2}(1+\sigma)\ln%
(1+\sigma)-\frac{1}{2}(1-\sigma)\ln(1-\sigma)\right]\nonumber\\%
&&-\frac{1}{2\beta}\int\wp(D_{B})\ln\left[%
1+2e^{-\beta{D_{B}}}\cosh(\beta{J\sigma})\right]dD_{B}.%
\end{eqnarray}%
\else
\end{multicols}\vspace*{-3.5ex}{ \tiny \noindent
\begin{tabular}[t]{c|}
\parbox{0.493\hsize}{~} \\ \hline
\end{tabular}
} 
\begin{equation}
\Psi (\sigma ) =-\frac{1}{2\beta }\left[ \ln 2-\frac{1}{2}(1+\sigma )\ln
(1+\sigma )-\frac{1}{2}(1-\sigma )\ln (1-\sigma )\right]
-\frac{1}{2\beta }\int\wp (D_{B})\ln \left[
1+2e^{-\beta D_{B}}\cosh (\beta J\sigma )\right] dD_{B}.
\end{equation}
{\tiny\hspace*{\fill}\begin{tabular}[t]{|c}\hline
\parbox{0.49\hsize}{~}\\
\end{tabular}}\vspace*{-2.5ex}%
\begin{multicols}{2}\noindent
\fi
From the minimization of $\Psi(\sigma)$ 
with respect to $\sigma $, we have the $A$-sublattice
magnetization, 
\begin{equation}
\sigma =\tanh \left[ \beta J\int\wp (D_{B})\frac{%
2\sinh (\beta J\sigma )}{e^{\beta D_{B}}+2\cosh (\beta J\sigma )}dD_{B}%
\right] ,
\end{equation}
where the random variable $D_{B}$ satisfies the probability distribution 
in Eq.\ (%
\ref{smprob}). We can now calculate various expectation values. For example,
we have 
\begin{eqnarray}
Q &=& \int\wp (D_{B})\left\langle S_{B}^{2}\right\rangle
dD_{B} \nonumber \\ &=& \int\wp (D_{B})\frac{2\cosh (\beta J\sigma )}{%
e^{\beta D_{B}}+2\cosh (\beta J\sigma )}dD_{B}.
\end{eqnarray}

The critical line comes from the condition 
\begin{equation}
\left. \frac{\partial ^{2}\Psi }{\partial \sigma ^{2}}\right| _{\sigma
=0}=0\Longrightarrow e^{\Delta }=2\frac{(K-1)-\frac{1}{3}pK^{2}}{1-\frac{2}{3%
}pK^{2}},  \label{smcwc}
\end{equation}
where $\Delta =\beta D$ and $K=\beta J$. The thermodynamic stability of the
critical line depends on the sign of the fourth derivative of $\Psi (\sigma
) $ at $\sigma =0$. There is then the possibility of a tricritical point,
given by the additional condition 
\begin{equation}
\left. \frac{\partial ^{4}\Psi }{\partial \sigma ^{4}}\right| _{\sigma
=0}=0\Longrightarrow K^2=\frac{3+9p+\sqrt{9-186p+177p^{2}}}{8p}.
\label{smcwtri}
\end{equation}
The stability of the tricritical point is determined by 
\begin{equation}
\left. \frac{\partial ^{6}\Psi }{\partial \sigma ^{6}}\right| _{\sigma
=0}\geq 0\Longrightarrow p\leq p_{m}=0.04485\ldots .,
\end{equation}
which means that the tricritical behavior is suppressed by disorder
concentrations larger than approximately $4.5\%$.

In Fig.\ \ref{fig1}, we plot some $D\times T$ 
phase diagrams for a set of typical
values of the concentration $p$. In the uniform case ($p=0$), there is just
a tricritical point, $P_{t}$. For $0<p\leq p_{m}=0.04485\ldots $, a
tricritical point still exists (see Fig.\ \ref{fig1}, for $p=0.04$). 
However, at low
temperatures and sufficiently large values of $D$, there appears a
low-density ($Q\rightarrow p$ as $T\rightarrow 0$) ferromagnetic phase,
which we call ferro-II phase; at fixed values of $D$, an increase of
temperature induces a second-order transition from the ferro-II to the
paramagnetic phase. This transition is represented by a critical line that
meets the first-order line at a critical end point, $P_{ce}$. This critical
end point separates the first-order line into two distinct regions: (i) at
higher temperatures, there are transitions between the usual, high-density ($%
Q\rightarrow 1$ as $T\rightarrow 0$) ferromagnetic phase (ferro-I) and the
paramagnetic phase; (ii) at lower temperatures, the transitions are between
the ferro-I and ferro-II phases, the first-order boundary ending at a
finite-temperature critical point, $P_{cs}$.

For $p_{m}=0.04485\ldots <p<3/59=0.05084\ldots $, the tricritical point is
replaced by a critical end point and a simple critical point, separated by a
first-order transition line between the ferromagnetic phases (see inset in
Fig.\ \ref{fig1}, for $p=0.05$).

For $p\geq 3/59$, the critical line is fully stable (see Fig.\ \ref{fig1}, 
for $p=0.08$). However, for $p\lesssim 0.1$, there still exists a small
finite-temperature region where there are (first-order) transitions between
the ferromagnetic phases.

\section{Bethe--Peierls approximation}

To give an estimate of the effects of thermal fluctuations, which are not
accounted for by the Curie--Weiss calculations, we now resort to a standard
self-consistent Bethe--Peierls approximation. As the model is defined on a
bipartite lattice, we have to consider two distinct clusters of coordination 
$q$ (see Fig.\ \ref{fig2}). 
In one of them, which we call $A$, the central site is
occupied by a spin $\sigma =\frac{1}{2}$, 
connected to $q$ spins of the $S=1$ type.
In the other cluster, called $B$, there is a central $S=1$ spin surrounded
by $q$ spin-$\frac{1}{2}$ variables. 
According to the standard prescription of the
Bethe--Peierls approximation, we assume that the boundary spins of cluster $A
$ are under the action of an effective magnetic field $\tilde{h}_{B}$ and an
effective crystal field $\tilde{D}$, while the boundary spins of cluster $B$
are in an effective magnetic field $\tilde{h}_{A}$. 
The crystal field acting
on the central site of a cluster $B$ is a random variable $D_{B}$. We also
consider external magnetic fields, $h_{A}$ and $h_{B}$, acting on the
central sites of clusters $A$ and $B$, respectively.

For the sake of simplicity, we assume that the same effective crystal field
$\tilde{D}$ acts on all boundary spins of cluster $A$ (and impose
self-consistency between both thermal and disorder averages associated with
the two clusters). In a more refined approach, we might introduce different
effective crystal fields to mimic the extended disorder of real materials.
However, we expect that, in the absence of external magnetic fields, the
assumption of a single $\tilde{D}$ is reasonable, at least in the
paramagnetic phase, where there is no long-range order; in particular, we
expect that the paramagnetic phase boundaries obtained by the two 
approaches are equivalent within the Bethe--Peierls approximation. We will
see that this is indeed supported by the calculations.

The partition functions associated with the two clusters are given by 
\ifpreprintsty
\else
\end{multicols}\vspace*{-3.5ex}{ \tiny \noindent
\begin{tabular}[t]{c|}
\parbox{0.493\hsize}{~} \\ \hline
\end{tabular}
} \fi
\begin{equation}
Z_{A}=e^{\gamma _{A}}\left[ 1+2e^{-\tilde{\Delta}}\cosh (\tilde{\gamma}%
_{B}+K)\right] ^{q}+e^{-\gamma _{A}}\left[ 1+2e^{-\tilde{\Delta}}\cosh (%
\tilde{\gamma}_{B}-K)\right] ^{q}
\end{equation}
and 
\begin{equation}
Z_{B}=\left[ 2\cosh (\tilde{\gamma}_{A})\right] ^{q}+e^{-\Delta _{B}}\left\{
e^{\gamma _{B}}\left[ 2\cosh (\tilde{\gamma}_{A}+K)\right] ^{q}+e^{-\gamma
_{B}}\left[ 2\cosh (\tilde{\gamma}_{A}-K)\right] ^{q}\right\} ,
\end{equation}
\ifpreprintsty
\else
{\tiny\hspace*{\fill}\begin{tabular}[t]{|c}\hline
\parbox{0.49\hsize}{~}\\
\end{tabular}}\vspace*{-2.5ex}%
\begin{multicols}{2}
\noindent
\fi
where $\gamma =\beta h$, $\Delta =\beta D$ and $K=\beta J$. The effective
fields $\tilde{\gamma}_{A}$, $\tilde{\gamma}_{B}$ and $\tilde{\Delta}$ are
determined by the consistency equations 
\begin{equation}
\sigma =\left[ \left\langle \sigma _{j}\right\rangle \right] _{\text{av}}=%
\frac{\partial \ln Z_{A}}{\partial \gamma _{A}}=\frac{1}{q}\int \wp (D_{B})%
\frac{\partial \ln Z_{B}}{\partial \tilde{\gamma}_{A}}dD_{B},  \label{smcem}
\end{equation}

\begin{equation}
S=\left[ \left\langle S_{j}\right\rangle \right] _{\text{av}}=\frac{1}{q}%
\frac{\partial \ln Z_{A}}{\partial \tilde{\gamma}_{B}}=\int \wp (D_{B})\frac{%
\partial \ln Z_{B}}{\partial \gamma _{B}}dD_{B},  \label{smcen}
\end{equation}
and 
\begin{equation}
Q=\left[ \left\langle S_{j}^{2}\right\rangle \right] _{\text{av}}=-\frac{1}{q%
}\frac{\partial \ln Z_{A}}{\partial \tilde{\Delta}}=-\int \wp (D_{B})\frac{%
\partial \ln Z_{B}}{\partial \Delta _{B}}dD_{B},  \label{smceq}
\end{equation}
where $\left\langle \cdots \right\rangle $ and $\left[ \cdots \right] _{%
\text{av}}$ indicate thermal and disorder averages, respectively. We point
out that the introduction of the effective crystal field $\tilde{D}$ is
essential to achieve consistency between the equations for the two clusters.

In order to analyze the critical behavior, it is convenient to choose the
magnetization $\sigma $, the temperature $T$, and the external fields $h_{B}$
and $D_{B}$, as the independent thermodynamic variables. Thus, the external
field $h_{A}$ is written as a function of those variables, and the
second-order transitions in zero external field ($h_{A}=h_{B}=0$) are given
by 
\begin{equation}
\left. \frac{\partial \gamma _{A}}{\partial \sigma }\right| _{\sigma
=0}=0\Longrightarrow Q_{0}=\frac{1}{(q-1)^{2}\tanh ^{2}K},  \label{smbcp}
\end{equation}
where the derivative is taken for fixed values of the remaining independent
variables, and 
\begin{equation}
Q_{0}\equiv \left. Q\right| _{\sigma =0}=\int \wp (D_{B})\frac{2\cosh ^{q}K}{%
e^{\Delta _{B}}+2\cosh ^{q}K}dD_{B}.
\end{equation}
To calculate the derivative in Eq. (\ref{smbcp}), we take the
(implicit) derivative of the consistency equations with respect to $\sigma $%
, imposing the condition $\sigma =0$ and eliminating the derivatives
involving $S$, $Q$ and the effective fields. Also, for $\sigma =0$, we have $%
S=\tilde{\gamma}_{A}=\tilde{\gamma}_{B}=0$, since those variables are odd
functions of $\sigma $ for $h_{A}=h_{B}=0$. Thus, the consistency equation
for $Q$ leads to 
\begin{equation}
\left. \tilde{\Delta}\right| _{\sigma =0}=\ln \left( 2\frac{1-Q_{0}}{Q_{0}}%
\cosh K\right) ,
\end{equation}
and the final result is 
\begin{equation}
\left. \frac{\partial \tilde{\gamma}_{A}}{\partial \sigma }\right| _{\sigma
=0}=\frac{1+\left[ 2(q-1)-q^{2}\right] V_{0}
+(q-1)^{2}V_{0}^{2}}{1+(q-2)V_{0}+(q-1)^{2}V_{0}^{2}},
\label{derga}
\end{equation}
where $V_{0}=Q_{0}\tanh^{2}K$.
For $q=2$, Eq.\ (\ref{derga}) reproduces the exact one-dimensional 
expression for the $A$-sublattice susceptibility (see Appendix \ref{app1}). 
In fact, for $q=2$, it is not
difficult to check that we regain all the exact one-dimensional results.

It is easy to see that, in the uniform case, corresponding to $\wp
(D_{B})=\delta (D_{B}-D)$, the critical line is given by 
\begin{equation}
\Delta =\ln \left\{ 2\left( \cosh K\right) ^{q-2}\left[ q(q-2)\cosh
^{2}K-(q-1)^{2}\right] \right\} ,
\end{equation}
in agreement with the results from the Bethe-lattice \cite{dasilva91} and
the cluster-variational \cite{tucker99} calculations. 

Using the binary distribution in Eq.\ (\ref{smprob}), we obtain 
\begin{equation}
Q_{0}=p\frac{2\cosh ^{q}K}{1+2\cosh ^{q}K}+(1-p)\frac{2\cosh ^{q}K}{%
e^{\Delta }+2\cosh ^{q}K}.
\end{equation}
Therefore, the critical line is given by 
\begin{equation}
e^{\Delta}=2\frac{(1-p)-\phi(K)}{\phi(K)}\cosh^{q}K,
\label{smclp}
\end{equation}
where
\begin{equation}
\phi(K)=\frac{1}{(q-1)^{2}}\frac{\cosh ^{2}K}{\cosh ^{2}K-1}
-p\frac{2\cosh^{q}K}{1+2\cosh ^{q}K}.
\end{equation}
In the $T\rightarrow 0$ ($K\rightarrow \infty $) limit, we have 
\begin{equation}
e^{\Delta }\simeq \frac{e^{qK}}{2^{q-1}}\frac{(q-1)^{2}-1}{1-p(q-1)^{2}},
\end{equation}
which has a real solution for $\Delta $ if 
\begin{equation}
1-p(q-1)^{2}>0\Longrightarrow p<p_{cr}=\frac{1}{(q-1)^{2}}.  \label{smpc}
\end{equation}

This last result should be anticipated for a Bethe lattice, as we can see
from the following arguments. Consider a Cayley tree where the sites
belonging to every other shell, say those on odd-numbered shells, are
occupied with probability $p$, while the remaining sites are always
occupied. If $q$ is the coordination of the tree, the average number of
paths from the root (shell $0$) to the first shell is given by $p(q-1)$,
while we have $p(q-1)^{2}$ paths from there to the second shell. According
to this reasoning, the average number of paths from the root to the $(2n)$th
shell is given by $p^{n}(q-1)^{2n}$. In order to have at least one path to
the surface of the tree ($n\rightarrow \infty $), it is required that $%
p(q-1)^{2}\geq 1$, which is just the condition in Eq. (\ref{smpc}). This
result, together with the reproduction of the exact one-dimensional
solution, would suggest that the present treatment also gives exact results on
the Bethe lattice even in the presence of disorder. However, as remarked in
previous similar treatments,\cite{bell75,young76} this works for the
paramagnetic phase only, because only then it is correct to assume that all
boundary sites are under the action of the same (zero) effective field. The
existence of a percolating cluster, which we do not take into account in
this treatment, prevents this approximation from still giving correct
results for the ordered phases.

We now consider Eq. (\ref{smclp}), in the infinite coordination limit ($%
q\rightarrow \infty $, $K\rightarrow 0$, $qK=\tilde{K}$). We then have 
\begin{equation}
e^{\Delta }=2\frac{\left( \tilde{K}-1\right) -\frac{1}{3}p\tilde{K}^{2}}{1-%
\frac{2}{3}p\tilde{K}^{2}},
\end{equation}
which agrees with Eq. (\ref{smcwc}) for the Curie--Weiss version of the
model.

The tricritical points are determined by Eq. (\ref{smbcp}) supplemented by
the condition 
\[
\left. \frac{\partial ^{3}\gamma _{A}}{\partial \sigma ^{3}}\right| _{\sigma
=0}=0,
\]
which is equivalent to
\begin{equation}
\frac{2q^{2}-10q+6}{(q-1)^{5}\tanh ^{2}K}%
+3qW_{0}\tanh ^{2}K=\frac{(q-2)(q-3)}{(q-1)^{3}},  \label{smbptri}
\end{equation}
where $W_{0}$ is given by 
\begin{equation}
W_{0}=\int \wp (D_{B})\left( \frac{2\cosh ^{q}K}{e^{\Delta _{B}}+2\cosh ^{q}K%
}\right) ^{2}dD_{B}.
\end{equation}
The tricritical points are stable if 
\[
\left. \frac{\partial ^{5}\gamma _{A}}{\partial \sigma ^{5}}\right| _{\sigma
=0}>0.
\]
To calculate this derivative, we again take implicit derivatives of the
consistency equations (up to fifth order) with respect to $\sigma $, at $%
\sigma =0$, and eliminate all derivatives involving $S$, $Q$ and the
effective fields. Unlike the previous analysis, we have not been able to
obtain closed-form expressions for the stability condition of the
tricritical point, but it is not difficult to perform a number of numerical
calculations.

For the uniform model, we have $W_{0}=Q_{0}^{2}$. Therefore, Eq. (\ref
{smbptri}) takes the form 
\begin{equation}
\tanh K=\frac{1}{q-1}\sqrt{\frac{5q-3}{q-3}},
\end{equation}
which is again identical to the result obtained from the Bethe-lattice \cite
{dasilva91} and the cluster-variational \cite{tucker99} calculations. Notice
that this equation has real solutions only if $q>4.561553\ldots $. Thus, the
Bethe--Peierls approximation does not predict a tricritical point for the
square lattice ($q=4$).

For the binary distribution in Eq.\ (\ref{smprob}), we have 
\begin{equation}
W_{0}=Q_{0}^{2}\left[ 1+\frac{p}{1-p}\left( 1-\frac{1}{Q_{0}}\frac{2\cosh
^{q}K}{1+2\cosh ^{q}K}\right) ^{2}\right] .
\end{equation}
In the infinite-coordination limit we can write 
\begin{equation}
W_{0}=\frac{1}{\tilde{K}^{4}}\left[ 1+\frac{p}{1-p}\left( 1-\frac{2}{3}%
\tilde{K}^{2}\right) ^{2}\right] ,
\end{equation}
which leads to the equation 
\begin{equation}
\tilde{K}^{2}-3\left[ 1+\frac{p}{1-p}\left( 1-\frac{4}{3}\tilde{K}^{2}+\frac{%
4}{9}\tilde{K}^{4}\right) \right] -2=0,
\end{equation}
at the tricritical point. Indeed, one of the solutions of this equation
corresponds to Eq. (\ref{smcwtri}), for the Curie--Weiss version of the
model, while the other solution represents a thermodynamically unstable
situation.

In Table \ref{tab1}, 
for various values of the coordination number $q$, and using the
binary distribution given by Eq. (\ref{smprob}), we give the corresponding
values of the concentration $p_{m}$, at which the tricritical point becomes
unstable, and the critical percolation concentration $p_{cr}$. We see that,
for $q\leq 10$, the tricritical behavior is suppressed for $p_{m}<p_{cr}$,
while, for $q\geq 11$, that suppression occurs for $p_{m}>p_{cr}$. As shown
in Table \ref{tab1}, 
$p_{m}$ increases with $q$, which indicates that disorder is
more effective for small coordination numbers.

As the effects of binary disorder strongly depend on the coordination, we
now discuss the phase diagrams for the typical cases. 

For $q=3$ and $4$, there are no tricritical points. The $D\times T$ phase
diagram displays just a fully stable critical line. The main effect of
disorder is to make the paramagnetic phase unstable at $T=0$, regardless of
the value of $D$, for $p$ larger than the critical percolation concentration 
$p_{cr}$. The phase diagrams in Fig.\ \ref{fig3}, for $q=3$, are in qualitative
agreement with the exact results for the honeycomb lattice (which is also
three-coordinated) under annealed disorder.\cite{goncalves92} At $T=0$,
there is even quantitative agreement with the value of the critical crystal
field at $p_{cr}$, given by $D_{cr}=5J/3$, although of course this agreement
does not extend to the value of $p_{cr}$ itself. Our results for $q=3$
and $q=4$ are also in qualitative agreement with those obtained by a 
real-space renormalization-group approach for the two-dimensional 
Blume-Emery-Griffiths model in a random crystal field. \cite{branco99}

For $5\leq q\leq 10$, the concentration $p_{m}$ above which the tricritical
point becomes unstable is lower than $p_{cr}$. For $p<p_{m}$, disorder
depressses the tricritical temperature, and shortens the first-order
transition line. For $p_{m}<p<p_{cr}$, the tricritical point is replaced by
a critical end point, $P_{ce}$, and a simple critical point, $P_{cs}$, as in
the Curie--Weiss version of the model. However, the paramagnetic phase is
stable at $T=0$ if $D>qJ$, and the first-order line reaches $D=qJ$ at $T=0$.
As $p$ increases, first the critical end point $P_{ce}$ and then the simple
critical point $P_{cs}$ reach the $T=0$ axis, at values of $p$ which can be
determined by a low-temperature expansion of the consistency equations (see
Appendix \ref{app2}). 
In Fig.\ \ref{fig4}, we plot the $D\times T$ phase diagram for $q=6$
and $p=0.011$. To determine the first-order lines shown in that figure, we
numerically solve the consistency equations to obtain the conditions $%
h_{A}\left( \sigma _{1}\right) =h_{A}\left( \sigma _{2}\right) =0$ and 
\begin{equation}
\int\limits_{\sigma _{1}}^{\sigma _{2}}h_{A}\left( \sigma \right) d\sigma =0,
\end{equation}
which correspond to a Maxwell construction.

For $q\geq 11$, we have $p_{m}>p_{cr}$, so the behavior of the system is
quite similar to the predictions of the Curie--Weiss version of the model.

\section{Conclusions}

We performed detailed calculations for the phase diagram of a
random-anisotropy mixed-spin Ising model both in the mean-field limit
(Curie--Weiss version of the model), in which thermal fluctuations are
neglected, and according to a standard self-consistent Bethe--Peierls
approximation (which turns out to be exact in one dimension). For a binary
distribution of crystal fields, we obtained closed-form expressions for the
critical lines and the location of the tricritical points. Depending on the
concentration $p$, the mean-field results for the $D\times T$ phase diagrams
predict novel first-order lines and multicritical points (besides a
ferromagnetic region, for all values of the crystal field, extending down to
the lowest temperatures). The Bethe--Peierls approximation shows that this
additional ferromagnetic region is suppressed for concentrations below a
certain percolation threshold. Also, the Bethe--Peierls results point out to
the absence of a tricritical behavior for lattices with coordination $q\leq 4
$. All results reported in this paper are in agreement with general
predictions for the effects of disorder on first-order transitions and
multicritical points (for a recent review, see a paper by Cardy 
\cite{cardy99}).

\acknowledgments{We thank T. A. S. Haddad for useful discussions.
This work was partially financed by the Brazilian agencies FAPESP and CNPq.}

\appendix
\section{Exact solution in one-dimension}
\label{app1}

For an open chain with $N+1$ sites ($N$ even), and in zero external field,
the Hamiltonian of the mixed-spin Ising model can be written as 
\begin{equation}
H=-J\sum_{j=1}^{N/2}\left( \sigma _{j}S_{j}+S_{j}\sigma _{j+1}\right)
+\sum_{j=1}^{N/2}D_{j}S_{j}^{2}.
\end{equation}
Given a disorder configuration $\{D\}=\{D_{1},\ldots ,D_{N/2}\}$,
we perform a partial trace over the spin variables $\{S_{j}\}$ to write 
\begin{eqnarray}
Z\{D\}&=&\sum_{\{\sigma \}}\sum_{\{S\}}e^{-\beta H} \nonumber \\
&=&\sum_{\{\sigma
\}}\prod_{j=1}^{N/2}\left\{ 1+2e^{-\Delta _{j}}\cosh [K(\sigma _{j}+\sigma
_{j+1})]\right\} ,
\end{eqnarray}
where $K=\beta J$ and $\Delta _{j}=\beta D_{j}$. Introducing a prefactor $%
A_{j}$, 
\begin{equation}
A_{j}^{2}=\left( 1+2e^{-\Delta _{j}}\right) \left[ 1+2e^{-\Delta _{j}}\cosh
(2K)\right] 
\end{equation}
and an effective interaction $\tilde{K}_{j}$, such that 
\begin{equation}
e^{2\tilde{K}_{j}}=\frac{1+2e^{-\Delta _{j}}\cosh (2K)}{1+2e^{-\Delta _{j}}},
\end{equation}
the partition function can be written as the factorized form 
\begin{eqnarray}
Z\{D\}&=&\sum_{\{\sigma \}}\prod_{j=1}^{N/2}A_{j}e^{\tilde{K}_{j}\sigma
_{j}\sigma _{j+1}} \nonumber \\
&=&\prod_{j=1}^{N/2}2\left[ 1+2e^{-\Delta _{j}}\cosh ^{2}K%
\right] .  \label{smz1d}
\end{eqnarray}

From Eq. (\ref{smz1d}), we obtain the thermal average 
\begin{equation}
\left\langle S_{j}^{2}\right\rangle _{\{D\}}=-\frac{\partial \ln Z}{\partial
\Delta _{j}}=\frac{2e^{-\Delta _{j}}\cosh ^{2}K}{1+2e^{-\Delta _{j}}\cosh
^{2}K},
\end{equation}
which depends on the value of the crystal field on the $j$th site only.
Since we are considering a nearest-neighbor one-dimensional model in zero
field, the thermal averages $\left\langle S_{j}\right\rangle $ and $%
\left\langle \sigma _{j}\right\rangle $ are zero. Performing the disorder
average, we obtain the expectation value 
\begin{equation}
Q=\int \left\langle S_{j}^{2}\right\rangle \prod_{i=1}^{N/2}\wp
(D_{i})dD_{i}=\int\wp (D_{j})\left\langle
S_{j}^{2}\right\rangle _{\{D\}}dD_{j}.  \label{smQ1d}
\end{equation}

For a given disorder configuration, the magnetic susceptibilities of the $%
\sigma $ and $S$ sublattices are given by 
\begin{equation}
\chi _{\sigma }\{D\}=\frac{1}{T}\lim_{N\rightarrow \infty }\frac{2}{N+2}%
\sum_{j=1}^{\frac{N}{2}+1}\sum_{k=1}^{\frac{N}{2}+1}\left\langle \sigma
_{j}\sigma _{k}\right\rangle _{\{D\}}
\end{equation}
and 
\begin{equation}
\chi _{s}\{D\}=\frac{1}{T}\lim_{N\rightarrow \infty }\frac{2}{N}%
\sum_{j=1}^{N/2}\sum_{k=1}^{N/2}\left\langle S_{j}S_{k}\right\rangle
_{\{D\}}.
\end{equation}
The two-spin correlation functions, 
\begin{equation}
\left\langle \sigma _{j}\sigma _{k}\right\rangle _{\{D\}}=\frac{1}{Z\{D\}}%
\sum_{\{\sigma \}}\sum_{\{S\}}\sigma _{j}\sigma _{k}e^{-\beta H}
\end{equation}
and 
\begin{equation}
\left\langle S_{j}S_{k}\right\rangle _{\{D\}}=\frac{1}{Z\{D\}}\sum_{\{\sigma
\}}\sum_{\{S\}}S_{j}S_{k}e^{-\beta H},
\end{equation}
can be calculated if we introduce the transformation 
\begin{equation}
\tau _{j}=\sigma _{j}\sigma _{j+1}\quad \text{with}\quad \tau _{0}=\sigma
_{1}.
\end{equation}
After some algebraic manipulations, for $j<k$, we have 
\begin{equation}
\left\langle \sigma _{j}\sigma _{k}\right\rangle _{\{D\}}=\prod_{i=j}^{k-1}%
\frac{2\sinh ^{2}K}{e^{\Delta _{i}}+2\cosh ^{2}K}
\end{equation}
and 
\begin{eqnarray}
\left\langle S_{j}S_{k}\right\rangle _{\{D\}}&=&\left( \prod_{i=j,k}\frac{%
\sinh 2K}{e^{\Delta _{i}}+2\cosh ^{2}K}\right) \nonumber \\
&&\times \prod_{i=j+1}^{k-1}\frac{%
2\sinh ^{2}K}{e^{\Delta _{i}}+2\cosh ^{2}K},
\end{eqnarray}
from which we obtain the expectation values 
\begin{eqnarray}
g_{\sigma }(\left| k-j\right| )&=&\int \left\langle \sigma _{j}\sigma
_{k}\right\rangle _{\{D\}}\prod_{i=1}^{N/2}\wp
(D_{i})dD_{i} \nonumber \\
&=&\left( Q\tanh ^{2}K\right) ^{\left| k-j\right| }
\end{eqnarray}
and 
\begin{eqnarray}
g_{s}(\left| k-j\right| )&=&\int \left\langle S_{j}S_{k}\right\rangle
_{\{D\}}\prod_{i=1}^{N/2}\wp (D_{i})dD_{i} \nonumber \\
&=& Q\left( Q\tanh
^{2}K\right) ^{\left| k-j\right| },
\end{eqnarray}
which depend on the distance between sites $j$ and $k$. The expectation
values of the susceptibilities are given by 
\begin{equation}
\left[ \chi _{\sigma }\right] _{\text{av}}=\frac{1}{T}\left[
1+2\sum_{r=1}^{\infty }g_{\sigma }\left( r\right) \right] =\frac{1}{T}\frac{%
1+Q\tanh ^{2}K}{1-Q\tanh ^{2}K}
\end{equation}
and 
\begin{equation}
\left[ \chi _{s}\right] _{\text{av}}=\frac{1}{T}\left[ Q+2\sum_{r=1}^{\infty
}g_{s}\left( r\right) \right] =\frac{Q}{T}\frac{1+Q\tanh ^{2}K}{1-Q\tanh
^{2}K},
\end{equation}
where $Q$ is determined by Eq. (\ref{smQ1d}).

\section{Low-temperature expansion}
\label{app2}

For the binary distribution, in the low-temperature limit ($K=\beta J\gg 1$%
), if we neglect terms of order $\exp \left( -2K\right) $ and higher, the
consistency equations (\ref{smcem})-(\ref{smceq}) for cluster $A$ lead to
the expressions 
\begin{equation}
\gamma _{A}=\frac{1}{2}\ln \frac{1+\sigma }{1-\sigma }+\frac{q}{2}\ln \frac{%
1-C_{+}}{1-C_{-}},  \label{ltga}
\end{equation}
\begin{equation}
S=\frac{1+\sigma }{2}C_{+}-\frac{1-\sigma }{2}C_{-},  \label{ltsa}
\end{equation}
and 
\begin{equation}
Q=\frac{1+\sigma }{2}C_{+}+\frac{1-\sigma }{2}C_{-},  \label{ltQa}
\end{equation}
where 
\begin{equation}
C_{\pm }=\frac{e^{\pm \tilde{\gamma}_{B}}}{e^{\tilde{\Delta}-K}+e^{\pm 
\tilde{\gamma}_{B}}}.
\end{equation}
For cluster $B$, we have 
\begin{eqnarray}
\sigma &=& p\tanh \left( q\tilde{\gamma}_{A}\right) \nonumber \\
&&+(1-p)\frac{\tau (\tilde{%
\gamma}_{A})\tanh (\tilde{\gamma}_{A})+\delta \tanh (q\tilde{\gamma}_{A})}{%
\tau (\tilde{\gamma}_{A})+\delta },  \label{ltma}
\end{eqnarray}

\begin{equation}
S=p\tanh \left( q\tilde{\gamma}_{A}\right) +(1-p)\frac{\delta }{\tau (\tilde{%
\gamma}_{A})+\delta }\tanh \left( \tilde{\gamma}_{A}\right) ,  \label{ltmb}
\end{equation}
\begin{equation}
Q=p+(1-p)\frac{\delta }{\tau (\tilde{\gamma}_{A})+\delta },  \label{ltQb}
\end{equation}
where 
\begin{equation}
\delta =\exp (qK-\Delta ),
\end{equation}
and 
\begin{equation}
\tau (x)=\frac{2^{q}}{\left( 1+\tanh x\right) ^{q}+\left( 1-\tanh x\right)
^{q}}.
\end{equation}

Solving Eqs. (\ref{ltsa}) and (\ref{ltQa}) for $C_{\pm }$ in terms of $%
\sigma $, $S$ and $Q$, and using Eqs. (\ref{ltma})-(\ref{ltQb}), we can
write Eq. (\ref{ltga}) in the form 
\begin{equation}
\gamma _{A}(\sigma )=\frac{1-q}{2}\ln \frac{1+\sigma }{1-\sigma }+q\tilde{%
\gamma}_{A}(\sigma ),  \label{ltstate}
\end{equation}
where $\tilde{\gamma}_{A}(\sigma )$ is determined from the solution of Eq. (%
\ref{ltma}). Notice that, according to Eqs. (\ref{ltstate}) and (\ref{ltma}%
), $\gamma _{A}(\sigma )$ and $\tilde{\gamma}_{A}(\sigma )$ depend on the
temperature through the parameter $\delta $ only. As $T\rightarrow 0$, this
parameter goes to zero (if $D>qJ$), or infinity (if $D<qJ$), except in the
vicinity of the point $P_{0}$ with coordinates $D=qJ$, and $T=0$, where $%
\delta $ can assume any value.

Since the equation of state (\ref{ltstate}) becomes asymptotically exact as $%
T\rightarrow 0$, it can be used to determine the values of $p$ at which the
critical end point and the simple critical point reach $P_{0}$, and thus
disappear. To do that calculation, we impose the conditions 
\begin{equation}
\gamma _{A}(\sigma _{e})=\left. \frac{\partial \gamma _{A}}{\partial \sigma }%
\right| _{\sigma =\sigma _{e}}=\left. \frac{\partial ^{2}\gamma _{A}}{%
\partial \sigma ^{2}}\right| _{\sigma =\sigma _{e}}=0,
\end{equation}
from which we obtain the values of $\sigma _{e}$, $\delta _{e}$ and $p_{e}$
at which the critical end point reaches $P_{0}$, and the conditions 
\begin{equation}
\gamma _{A}(\sigma _{s})=\left. \frac{\partial \gamma _{A}}{\partial \sigma }%
\right| _{\sigma =\sigma _{s}}=\int_{0}^{\sigma _{s}}\gamma _{A}(\sigma
)d\sigma =0,
\end{equation}
which give the corresponding values $\sigma _{s}$, $\delta _{s}$ and $p_{s}$
for the simple critical point.

\begin{figure}
\begin{center}
\includegraphics[totalheight=7.8cm,angle=-90]{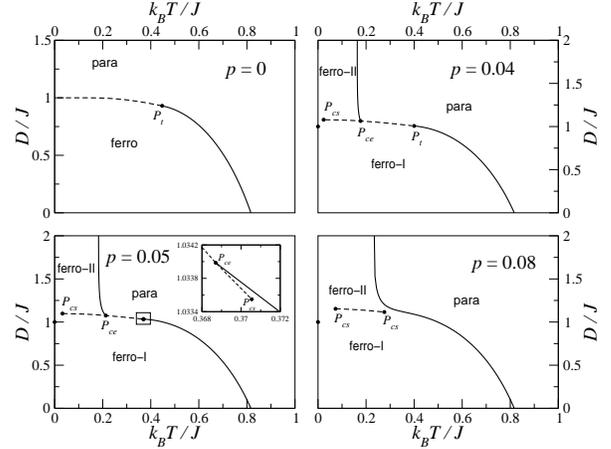}
\end{center}
\caption{Phase diagrams of the Curie--Weiss version for typical values
of the disorder concentration $p$.}
\label{fig1}
\end{figure}

\begin{figure}
\begin{center}
\includegraphics[width=7.5cm]{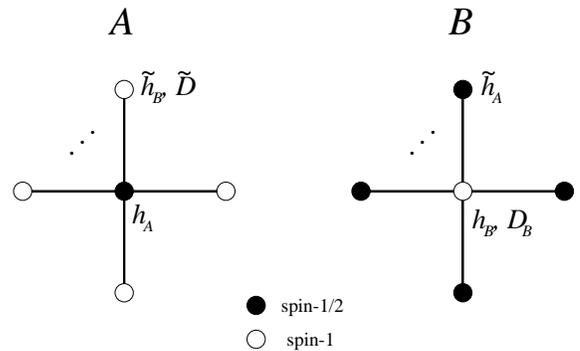}
\end{center}
\caption{Clusters used in the Bethe--Peierls approximation.}
\label{fig2}
\end{figure}

\begin{figure}
\begin{center}
\includegraphics[totalheight=7.5cm,angle=-90]{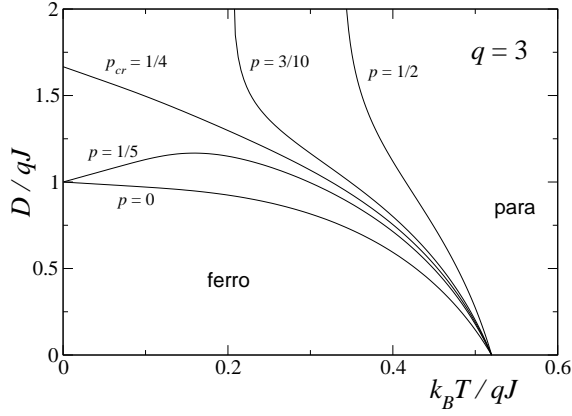}
\end{center}
\caption{Phase diagrams for coordination $q=3$ according to the Bethe--Peierls
approximation.}
\label{fig3}
\end{figure}

\begin{figure}
\begin{center}
\includegraphics[totalheight=7.5cm,angle=-90]{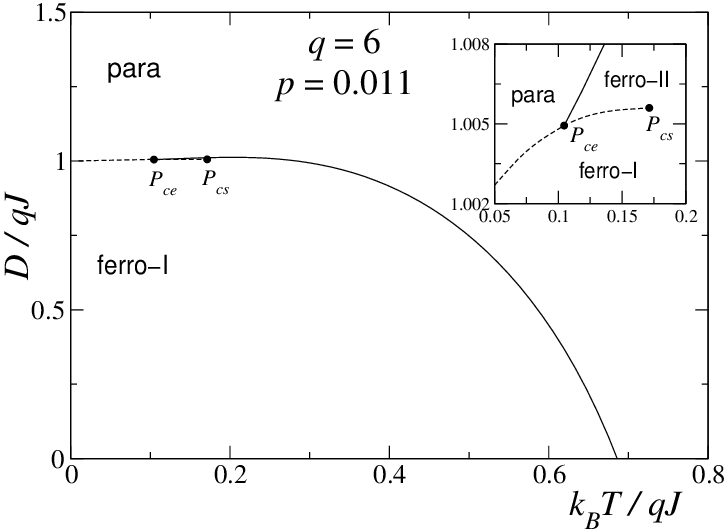}
\end{center}
\caption{Phase diagram for coordination $q=6$ and disorder concentration
$p=0.011$ according to the Bethe--Peierls approximation.}
\label{fig4}
\end{figure}

\begin{table}
\caption{Values of the critical percolation concentration $p_{cr}$ and the
concentration $p_{m}$ at which the tricritical point becomes unstable, as
functions of the coordination $q$ in the Bethe--Peierls approximation.}
{\centering \begin{tabular}{ccc} 
\( q \)&
\( p_{cr} \)&
\( p_{m} \)\\
\tableline 
5&
\( 6.25\times 10^{-2} \)&
\( 7.4161\times 10^{-4} \)\\
6&
\( 4\times 10^{-2} \)&
\( 2.0454\times 10^{-3} \)\\
10&
\( 1.2346\times 10^{-2} \)&
\( 9.8265\times 10^{-3} \)\\
11&
\( 1\times 10^{-2} \)&
\( 1.1665\times 10^{-2} \)\\
20&
\( 2.7701\times 10^{-3} \)&
\( 2.3001\times 10^{-2} \)\\ 
100&
\( 1.0203\times 10^{-4} \)&
\( 3.9707\times 10^{-2} \)\\
\( \infty  \)&
\( 0 \)&
\( 4.4850\times 10^{-2} \)\\
\end{tabular}\par}\vspace{0.3cm}
\label{tab1}
\end{table}

\ifpreprintsty
\else
\end{multicols}
\fi

\end{document}